# Robustness to Model Uncertainties Drives More Rapid $CO_2$ Emissions Reductions

Lisa Rennels[1], Frank Errickson[2], David Smith[3], Bryan Parthum[3], Klaus Keller[4], David Anthoff[5]

[1] Doerr School of Sustainability, Stanford University; Stanford, CA 94301, USA

[2] School of Public and International Affairs, Princeton University; Princeton, NJ 08540, USA

[3] Environmental Protection Agency; Washington, DC 20004, USA

[4] Thayer School of Engineering; Dartmouth College; Hanover, NH 03755, USA

[5] Energy and Resources Group, University of California Berkeley; Berkeley, CA 94720, USA

## Abstract

Evaluating the economic impacts of climate policies is important for designing a response to climate change. One typical approach to assessing mitigation policy options uses integrated climate-economy models to analyze tradeoffs between the costs of reducing greenhouse gas emissions and the benefits of reducing climate damages. However, the uncertainty characterizing these models poses significant challenges for policymakers. We address this difficulty using a robust decision-making framework to evaluate mitigation policy. We show that a shift from a decision framework that maximizes expected outcomes to one that is averse to regret suggests more aggressive emissions reductions. Uncertainties about socioeconomic trajectories and the magnitude and functional form of climate damages create the asymmetric consequences of weak mitigation policy that encourage aggressive emissions reductions and precaution in the face of uncertainty.

## Main Text

Anthropogenic climate change drives sizable risks to human and natural systems[1]. International agreements call for mitigating greenhouse gas emissions to net zero by mid-century[2,3]. However, designing such a mitigation strategy is a nontrivial task. For example, in some regulatory contexts United States policy requires quantification of the costs of mitigation alongside the benefits of avoided damages from climate change[4]. To do this, researchers use a variety of analytical frameworks and integrated assessment models (IAMs) to compare the relative strengths and weaknesses of mitigation policy options. IAMs can be used to evaluate policy options through a calculation of the net benefits and costs of a proposed policy or by computing the social cost of carbon[5-9].Within an optimization framework, these models can also generate carbon dioxide mitigation pathways to meet policy objectives, such as maximizing expected net benefits over time or meeting a policy target like keeping global temperatures below 2°C.

A fundamental difficulty faced by policymakers is consolidating results across many plausible models–all with unique parameterizations and structures. In other words, IAMs are plagued by "deep uncertainty", defined as a "situation where the system model and the input parameters to the system model are not known or widely agreed on by the stakeholders to the decision"[10]. Parallel work and literatures also refer to deep uncertainty as Knightian uncertainty, a term from the foundational literature separating known and unknown probabilities of outcomes[11], or ambiguity, formalizing Knightian uncertainty in the ambiguity aversion literature[12]. Despite decades of research, stakeholders and policymakers alike still face uncertainties about core structural elements including socioeconomic projections, abatement costs in an evolving technological environment, and the economic impacts of historically unprecedented temperature anomalies. Analysts can list a breadth of plausible future scenarios and relationships but are generally unable to assign them relative probabilities[13]. Assessments that explicitly account for these uncertainties are critical given the costs of inaction. Constructing policy against this backdrop is challenging[14], and a better representation of downside risk could yield different results[15].

Adopting a specific objective function, a single IAM structure, and single characterization of uncertainties, produces a single mitigation pathway. Since there are multiple IAM structures and multiple characterizations of uncertainties, policymakers face the complicated task of choosing or combining these results. One solution is to simply choose a single IAM based on some criteria and ignore the others. Alternatively, one may conclude that the uncertainty is too widespread and reject using integrated assessment models in climate policy analysis and, instead, prefer to rely on expert elicitation, simpler models, or to refocus efforts on cost-effectiveness analysis[16-18]. A third option applies the principle of insufficient reason[19, 20] by assigning equal probability to all considered model structures and taking the average of model results[7]. While this third approach seems attractive for its simplicity, it hinges on the assumption that we have no reason to believe one outcome is more likely than the other, which is not always the case for IAM structures[21].

Here, we take a different approach to decision-making under uncertainty and adopt a robust decision-making method that directly samples the deep structural uncertainty of integrated assessment models. Robust decision-making methods provide a framework for using multiple IAMs to inform climate policy discussion without requiring explicit probabilities for each structure[22]. This approach accommodates disaggregated results, presenting how different strategies perform across a range of futures and retaining information that an averaging approach loses. These disaggregated outputs enable explicit, transparent examination of tradeoffs and analysis of which elements of structural uncertainty are most consequential for the result.

A well-established body of literature surrounds the use of these methods for environmental problems and policy analysis, including water and energy resource management, coastal flood risk management, and climate change mitigation[23-27]. A subset of this literature focuses on integrated assessment modeling and climate policy, providing several key innovations. First, several studies outline the theoretical foundations for applying a single robustness approach, namely the minimizing maximum (minimax) regret criteria, to climate policy and IAMs. These include relatively stylized examples that vary the parameterization of a single model[28-30]. For example, one study presents robust decision-making as a useful framework to support climate policy and represents the uncertainty of a single integrated assessment model (DICE) using three discrete levels each for the climate sensitivity and damage function coefficient parameters[29]. Second, a few notable studies analyze a wider set of structures from published models and/or compare the results from multiple criteria for choosing the most robust policy[31-34]. For example, one study applies three different robustness criteria and characterizes climate uncertainty by substituting in four distinct temperature models, including three from foundational integrated assessment models: DICE, FUND, and PAGE[33]. Third, a recent study provides additional insights into how optimizing for different objectives, defined as reaching a climate target or limiting social inequality, reveal tradeoffs in climate policy[31].

This study adheres most closely to the approaches discussed in the previous paragraph. The literature contains several other approaches that are rooted in macroeconomic theory that examine climate-economy models and climate policy under conditions of Knightian uncertainty or ambiguity. Several studies provide empirical evidence that agents are ambiguity averse[12, 35]. There is a vigorous and ongoing debate whether to use this descriptive observation as a normative foundation for models of decision-making[36]. This literature also provides theoretical and explicitly parameterized frameworks that can represent ambiguity aversion, including smooth ambiguity aversion[37] and robust control[38]. This is an area of active and insightful research, expertly discussed in reviews such as those by Heal and Millner (2018)[36] and Jensen and Traeger (2024)[39] . Here we adopt a parsimonious approach and just focus on a relatively simple model-based robustness approach that examines the broad structural uncertainty of existing IAMs.

Guided by the literature, we hypothesize that adopting a robust decision-making framework and a more comprehensive treatment of structural uncertainty of IAMs may yield more stringent near-term mitigation policies. We incorporate both deep structural and quantifiable parametric uncertainties across a sample of 100 unique IAM structures. We find that adopting this robust decision-making framework, that is shifting from expected utility maximization to regret aversion, results in more rapid mitigation and more strictly limiting temperature increases than suggested by simply averaging models. The most robust policy exhibits full decarbonization by the year 2050 and limits the expected global mean temperature anomaly to an average of 2.2°C above pre-industrial across the considered possible future scenarios, or states of the world. This policy takes a precautionary approach, mitigating rapidly to avoid risking the potential consequences of a particularly high-emissions scenario combined with a high-damages scenario.

Our analysis encompasses both optimal policy generation and evaluation of each policy across many states of the world. This framework allows us to decompose results and derive insights into the dominant drivers of regret. We find that uncertainty in the temperature-damage relationship and in the future socioeconomic scenarios that project population growth, economic growth (as measured by gross domestic product (GDP)), and $CO_2$ emissions creates asymmetric regret from under-abatement. This asymmetry is the key driver of the precaution and aggressive mitigation suggested in our analysis. Our approach also allows us to directly compute and compare the optimal policy resulting from assigning equal probability to each model. Finally, we incorporate a wide range of structural uncertainties in addition to a comprehensive treatment of quantifiable parametric uncertainty through calculations of expected welfare and expected regret.

**Selecting the Most Robust Policy**

We create 100 unique model structures from the published literature, use these to derive mitigation policy candidates, employ the minimizing maximum regret (minimax) criterion to examine the performance of each policy across a range of possible future scenarios, and rank candidates by robustness (Fig. 1, Extended Data Table 1).

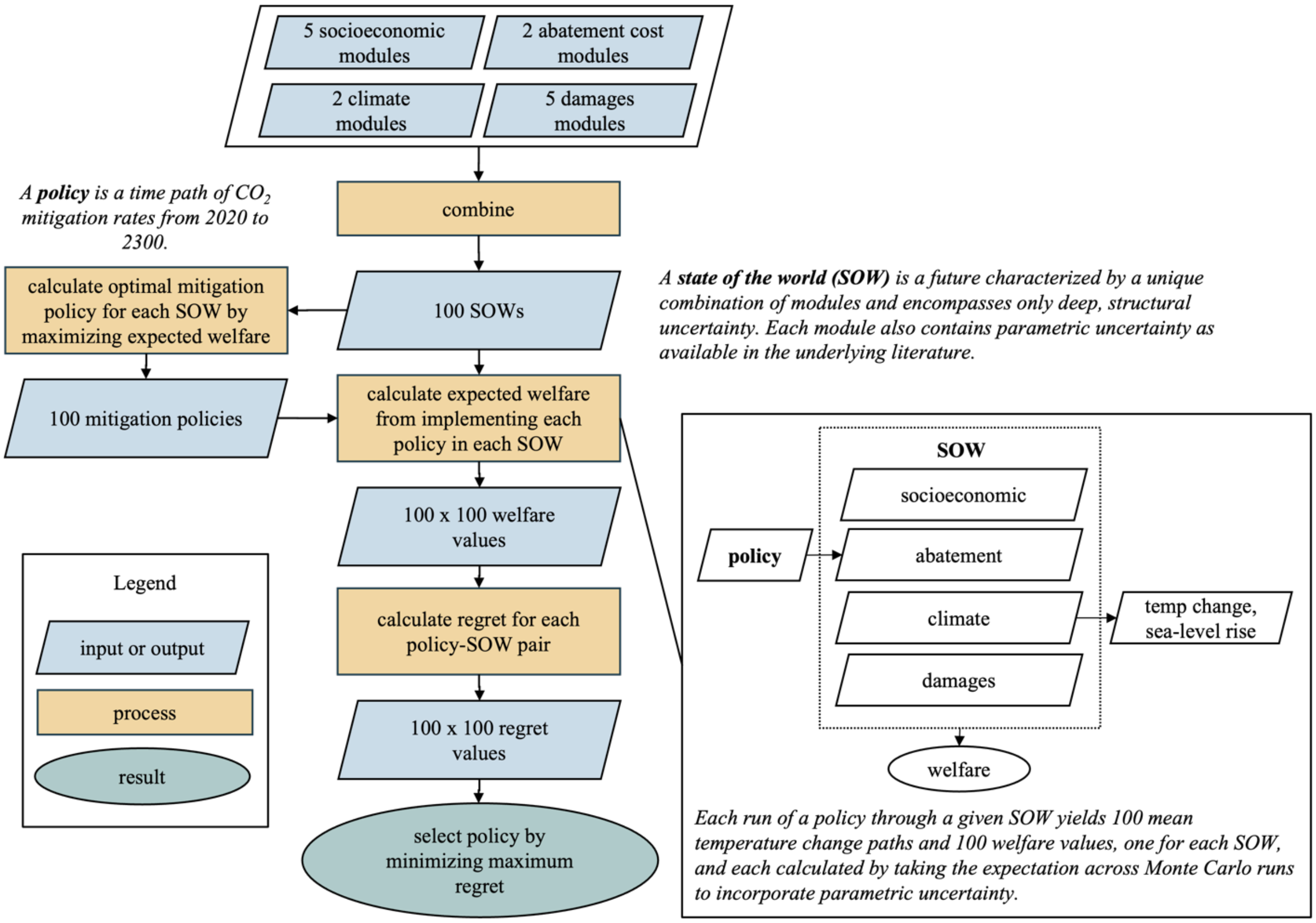

**Fig. 1. Methods Flowchart.** The high-level inputs, outputs, and processes of the methods. Blue trapezoids represent inputs or outputs, gold rectangles processes, and the green ovals the result.

We consider an IAM to be composed of four primary modules: socioeconomic, climate, abatement costs, and damages (Extended Data Table 2)[40]. We characterize the deep uncertainty of IAMs as structural uncertainty whereby we do not have a strong consensus on the probabilities of individual IAM components being the correct representation of a given module. Most published studies using IAMs account for quantifiable, or shallow, uncertainty through parameters with assigned probability densities[8, 41, 42]. We also include these parametric uncertainties in each module and incorporate them in each Monte Carlo simulation (See Methods).

IAMs can generate mitigation policy strategies. For each IAM, we find the path of mitigation rates that balance costs of abatement with damages of climate change and maximize expected welfare using a social welfare function (see Methods). Repeating this step for each model structure yields one optimal policy for each structure, generating a total of 100 candidate policies. The optimization generating each of these candidate policies samples across the shallow, parametric uncertainty within that structure. We then compute expected welfare across the parameterizations of a given future state of the world. This collapses parametric uncertainty using the expected value but retains the discrete enumeration of possible model structures under deep uncertainty. Intuitively, a policy will be the optimal choice if it maximizes welfare while assuming the model structure that accurately depicts the future.

We finally identify the policy that minimizes maximum regret. This criterion has a long history in the decision theory literature[43] and more recent applications in the climate policy literature[29, 32-34]. Regret theory differs from a classic expected utility framework in the following way. Instead of picking the strategy that has the highest expected value outcome, one instead asks "what happens if we are wrong?" and "how much better could we have done?" Making a decision under uncertainty can lead to a suboptimal outcome and this approach puts regret at the center of the analysis, where regret is defined as the difference between the realized suboptimal outcome and the best possible outcome[12, 43]. We focus on the minimax regret criterion to rank policies by robustness and explore how the deep structural uncertainty of IAMs affects the robustness of climate policy design. Following the classic definition[43] we calculate regret for a policy-SOW combination as the difference between how the policy performs in the SOW, and how the optimal policy for that SOW performs. The most robust policy under the minimax regret criterion is the one with the smallest maximum regret across the resulting 100 model structures and parameter values.

**Increasing Decarbonization Speed Increases Robustness**

The robust policy is at the upper end of decarbonization rates compared to the full set of candidate policies, reaching 100% mitigation in 2050 and rapidly incorporating negative emissions after that year (Fig. 2, Extended Data Fig. 1). Most policies decarbonize rapidly until 2030. After 2030, the rates begin to diverge with some policies continuing high rates of reduction and others leveling off. The earliest year of zero $CO_2$ emissions across candidate policies is 2040. However, about 80% of policies do not fully decarbonize until after 2050 and 30% until after 2100. Comparing the robust mitigation path with a few alternative policies can provide intuition about the implications of using the minimax regret method, and robust decision-making more broadly, to select preferred mitigation policies.

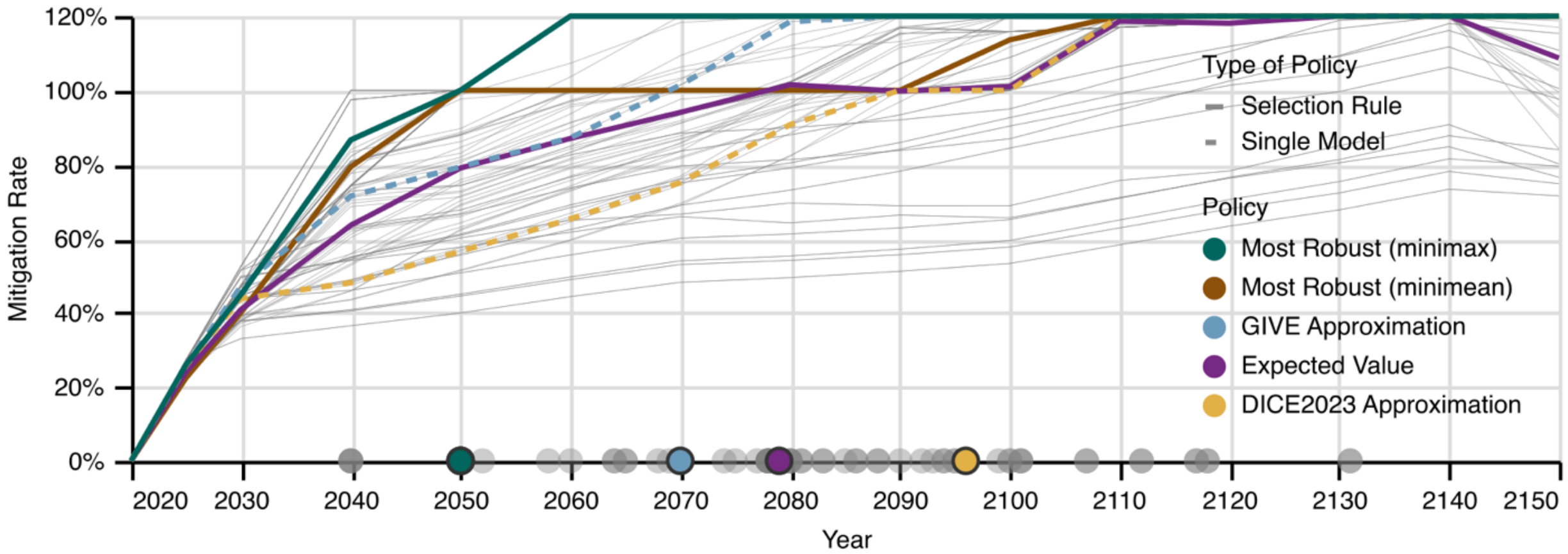

**Fig. 2. Optimal mitigation rate paths vary in robustness: the most robust path decarbonizes rapidly.** All considered candidate policies (grey lines) with five highlighted representative policies including (1) the most robust policy when minimizing maximum regret (green) (2) the most robust policy when minimizing mean regret (brown) (3) an additional policy, not considered in robustness analysis, that optimizes expected welfare across all states of the world (purple), and policy approximations for commonly cited integrated assessment models (4) the Greenhouse Gas Impact Value Estimator (GIVE) (blue) and (5) the Dynamic Integrated Model of the Climate and the Economy 2023 (DICE2023) (gold). Solid colored lines show policies selected based on a selection rule over a suite of models and dashed colored lines show policies corresponding to a single model. Each point along the x-axis shows the date when each policy reaches full decarbonization. Colored dots with black outlines show decarbonization dates for the five highlighted representative policies. Note that decarbonization dates for both robust policies are the same, and as such the dot for the minimean policy is under that of the minimax policy.

The robust policy is more stringent than the optimal policies for structural approximations of two popular IAMs, DICE2023[44] and GIVE[8] (See Methods for details on these approximations). Applying the principle of insufficient reason[19, 20] and taking each IAM as equally likely also generates a comparatively relaxed policy. Assuming there is no evidence that one structure is more or less likely than another, we assign each structure equal probability and maximize the expected welfare across all future states of the world to produce an additional policy not included in the primary set of 100 candidate policies. The resultant policy fully decarbonizes nearly thirty years later than the most robust policy and shows a slower rate of mitigation particularly in the second half of the century. This result suggests that policy recommendations based on averaging over the results of existing IAMs may lead to more cumulative emissions, and higher temperatures, than those derived from robust decision-making criteria.

The minimax regret criterion approximates a highly regret-averse attitude and is sensitive to worst-case outcomes[45]. An alternative robustness metric ranks policies by minimizing the mean regret (minimean). A robustness criteria based on the central tendency of regret, as opposed to the worst-case, is more tempered against worst cases. We find that the policy selected using the minimean criteria is nearly identical, albeit with slightly slower decarbonization, to that of the minimax criteria until 2050, when both fully decarbonize (Fig. 2). They then diverge, with the policy selected through the minimean criteria delaying net negative mitigation for several decades. Our findings are relatively generalizable to the regret framing, especially in the near-term, and not as sensitive to the form of regret.

**Increasing Robustness Reduces Risk of Extreme Warming**

Applying robustness criteria, especially highly regret averse ones like the minimax criteria, reduces the occurrence of extreme rise in global temperatures (Fig. 3, Extended Data Fig. 2). Temperature outcomes from the robust policy are varied, subject to the possibility of high-emissions future scenarios, but the set of outcomes is noticeably constrained and reduces the occurrence of extreme high warming outcomes. Strict decarbonization stabilizes temperatures more quickly, as opposed to allowing for larger increases past the end of the century.

Each policy produces 100 global mean temperature anomaly trajectories, one for each possible state of the world, each calculated by taking the expectation across Monte Carlo runs to incorporate parametric uncertainty. Across the full range of 10,000 sampled expected temperature paths generated by combining 100 policy candidates with 100 states of the world, maximum temperatures are on average 2.6°C above pre-industrial, but range from 1.8 to 5°C. In contrast, across just the subset of 100 temperature paths generated by the robust policy, maximum temperatures are on average 2.2°C, and from 1.8 to 2.6°C depending on the realized SOW (Extended Data Fig. 3, Extended Data Fig. 4). While the averages only differ by a fraction of a degree, the most robust path cuts off the upper tail of extreme warming. The most robust policy takes immediate and rapid action to stabilize temperatures by around 2100, as compared to many policies that reach peak temperatures well beyond 2100 (Fig. 3).

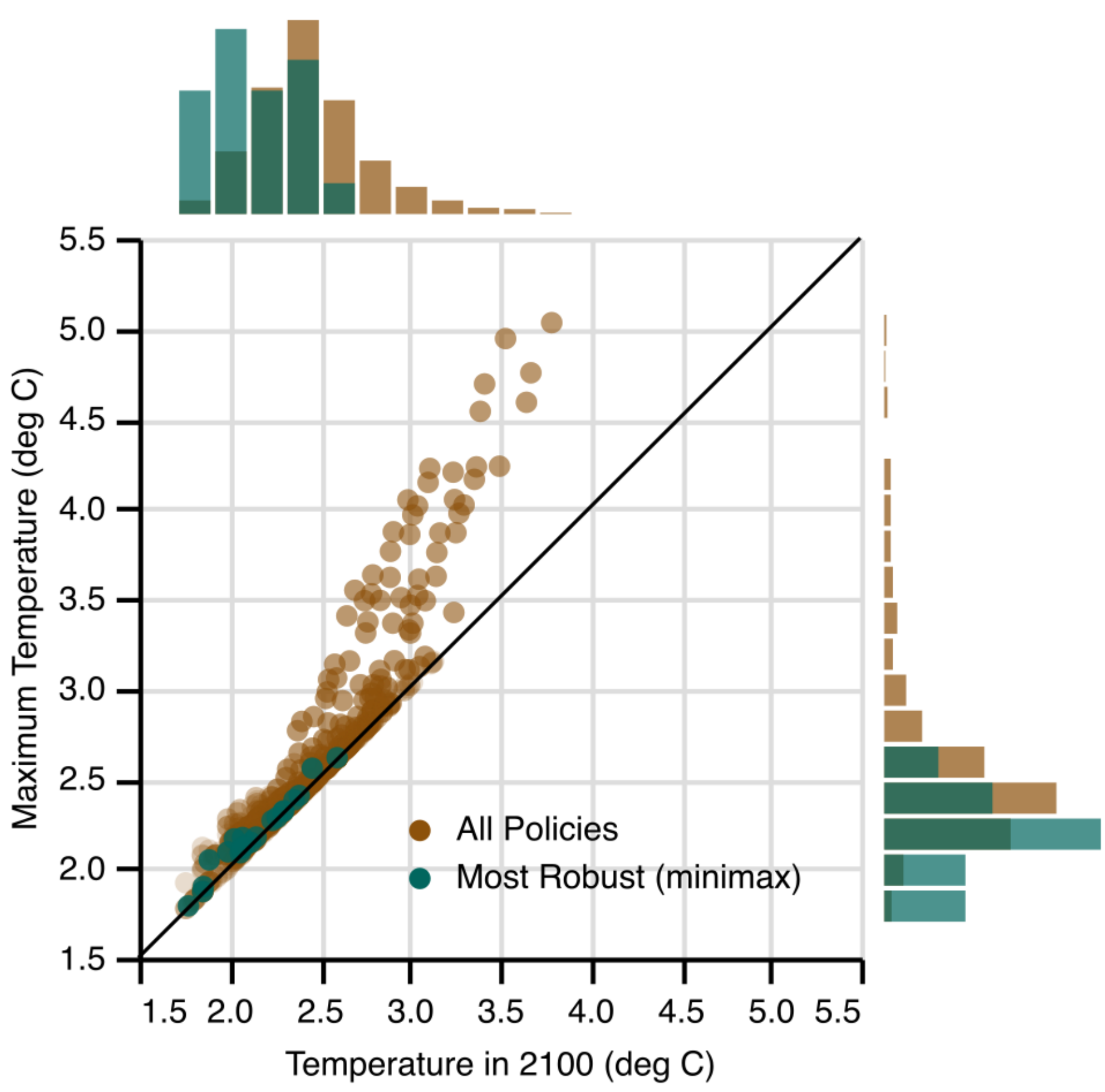

**Fig. 3. The robust policy limits temperature increase and avoids worst-case warming scenarios.** Points show the relationship between temperature in 2100 and the maximum expected temperature for all policies (green) and the most robust policy (brown). Normalized histograms show the distribution of mean temperature in 2100 (top, x-axis) and maximum mean temperature (right, y-axis) policy-state of the world pairs using the same color scheme.

**Asymmetric Regret from Under-abatement Motivates More Stringent Action**

The deep uncertainty about damage impacts and socioeconomic trajectories in future states of the world drives a noticeable asymmetry in regret patterns (Fig. 4). The possibility of severe economic damages with respect to temperature, such as those resulting from persistent decreases in economic growth, paired with uncertainty about future socioeconomic scenarios results in strongly negative consequences of slow mitigation policy (Fig. 5, Extended Data Fig. 5, Extended Data Fig. 6). The corresponding asymmetric regret from under-abating drives the precautionary tendency to abate aggressively in the face of uncertainty. This uneven relationship is especially pronounced for those impact functions that model an effect on economic growth, such as Burke et al. (2015)[46]. Previous studies recognize these two modules as both influential for conclusions about climate policy and lacking in scientific consensus, posing clear challenges for interpreting and using results[47-52].

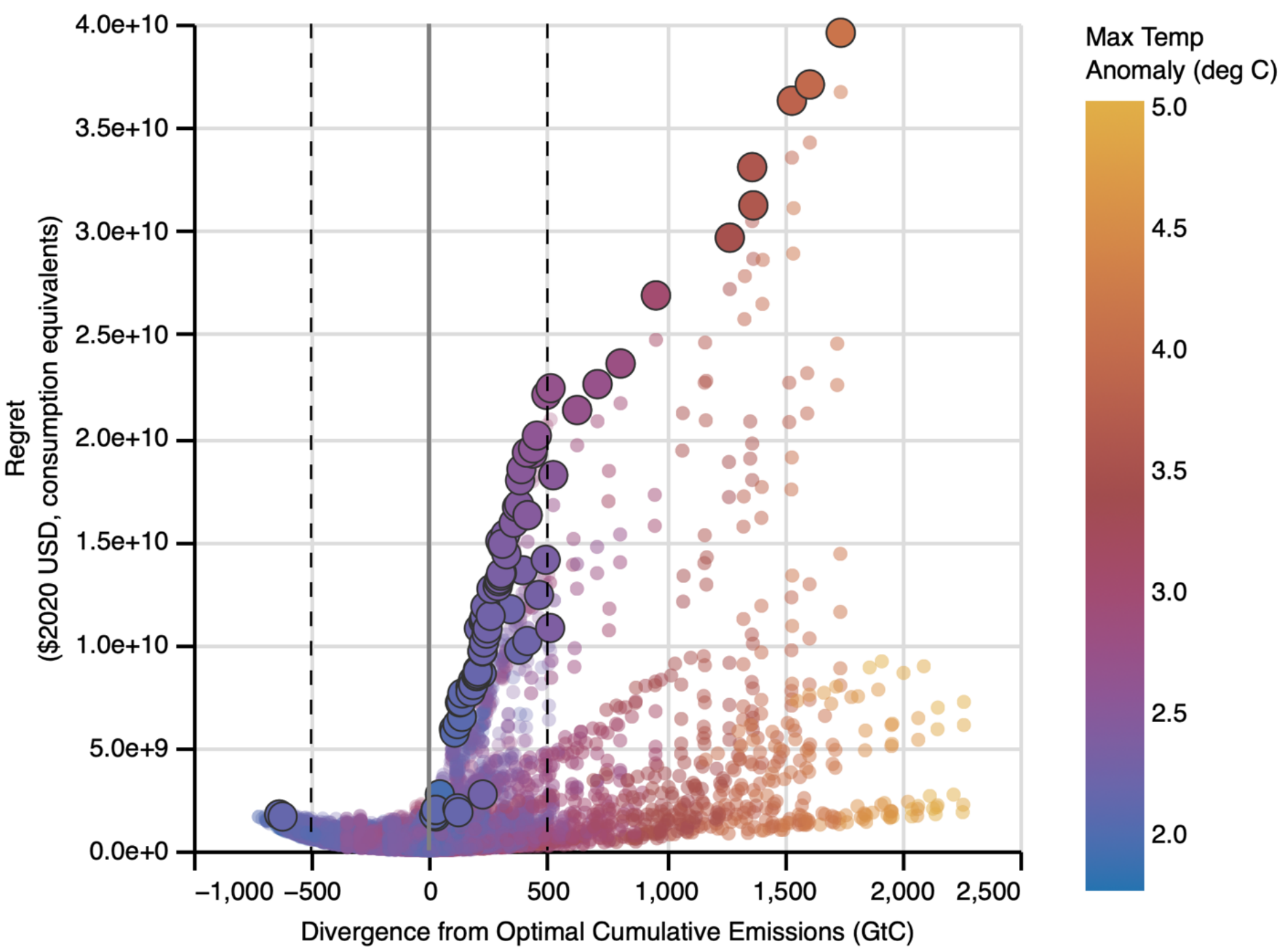

**Fig. 4. Asymmetric regret and the minimax criterion.** Each point represents a policy-state of the world pair, with the divergence from the optimal cumulative emissions in a realized state of the world on the x-axis and the subsequent regret, on the y-axis. Color indicates the maximum expected temperature anomaly along the realized temperature path. Larger circles with a black outline represent the pair that creates maximum regret, one for each policy. This is the regret value that determines the rank of a given policy under the minimax regret criterion.

Cumulative emissions are another useful summary statistic for policy candidates. While in previous sections we use the year a policy reaches net-zero emissions as a metric for comparisons, the time path of mitigation is highly consequential for temperature and resultant damages. This trajectory dictates cumulative $CO_2$ emissions, which determine the maximum warming for decades well-after emissions reduce to zero[53]. Persistent negative emissions may eventually reduce the temperature anomaly below that maximum[3, 53, 54].

We quantify suboptimality of a policy by the difference between optimal cumulative emissions in a future state of the world, and the cumulative emissions under the implemented policy. This metric clearly shows disproportionately high regret from emitting above the optimal amount as opposed to being overly-conservative (Fig. 4). For example, average regret from over-emitting ~ 500 GtC (+/- 10 GtC) is about four times as high as that for under-emitting the same amount. For the same deviation of 500 GtC (+/- 10 GtC), maximum regret across all policy-SOW pairs from over-emitting is about fifteen times as high as that for under-emitting.

Which uncertainty surrounding specific structural components drive the asymmetry? We find that the ray-like groups of points (Fig. 4), are largely determined by two characteristics of states of the world: the damage module and socioeconomic module (Fig. 5, Extended Data Fig. 5, Extended Data Fig. 6). The damage function realized in a future state of the world drives the regret asymmetry and creates the risk of extremely high welfare losses from under abatement. Future states of the world where climate damages have persistent impacts on economic growth dominate the highest regret policy-SOW pairs, marking a clear delineation from those functions which estimate damage to the economy in a given year that does not persist into the future (Fig. 5). This suggests that while experts continue to grapple with forecasting the channels of impacts climate change will have on the economy, a regret-based approach encourages hedging against the possibility of persistent impacts on economic growth[55]. Incorrect assumptions about future socioeconomic scenarios also have sizable consequences for future regrets (Extended Data Fig. 6). For example, optimistic predictions of future emissions such as those in Shared Socioeconomic Pathway 1 produce low mitigation rate policies with high consequences for welfare if these predictions fail.

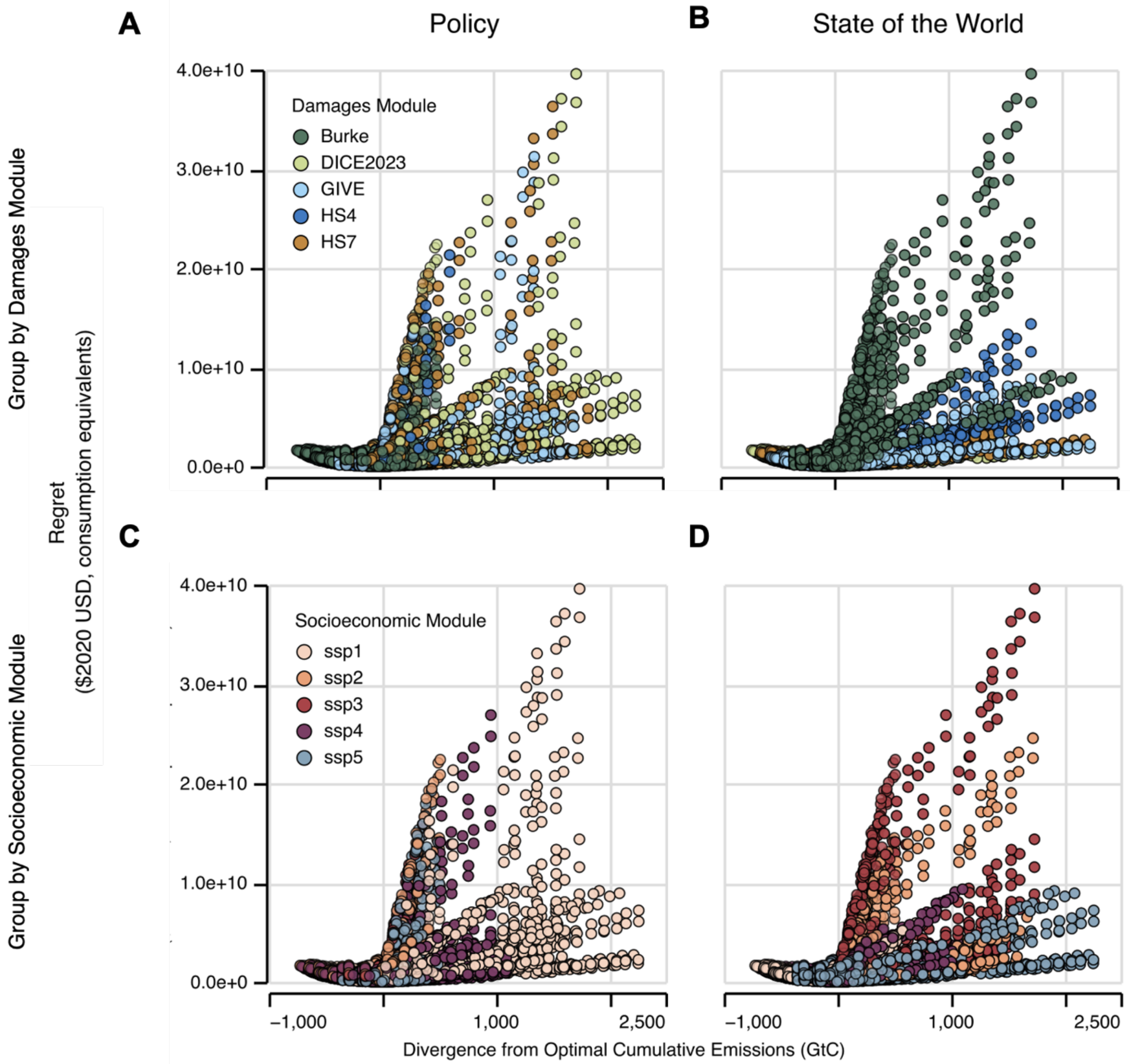

**Fig. 5. The socioeconomic and damages modules drive patterns of regret and asymmetry.** Each point represents a policy-state of the world pair, with the divergence from the optimal cumulative emissions in a realized SOW on the x-axis and the subsequent regret on the y-axis. Rows indicate module type, with the damages module on the top **(A, B)** showing the five damage modules Burke et al. (Burke), Dynamic Integrated Model of the Climate and the Economy 2023 (DICE2023), Greenhouse Gas Impact Value Estimator (GIVE), and Howard and Sterner specifications 4 and 7 (HS4, HS7), and the socioeconomic module on the bottom **(C, D)** showing the five benchmark Shared Socioeconomic Pathways (SSPs). Column 1 **(A, C)** groups the points by characteristic of the policy (e.g. the socioeconomic scenario assumed during the optimization that generated a given policy). Column 2 (**B, D**) groups the points by characteristic of the state of the world (e.g. the socioeconomic scenario realized to characterize a future).

## Discussion

We adopt an overall simple analytical framework to produce transparent and intuitive results. This simplicity results in caveats and points to research needs. For example, incorporation of additional elements of deep uncertainty, such as climate system tipping points, may enhance our understanding of the relative contributions of the breadth of uncertainties on the robustness of given climate policies. Similarly, adding abatement costs structures which incorporate social, political, and technological interactions to better model plausible trajectories would significantly enhance our analysis. The paucity of available abatement cost functions that can be readily plugged into our framework significantly limits our exploration of uncertainty in that module, especially in comparison to the damages module. What we are able to include is model uncertainty specifically over negative emissions, which is a consequential and highly uncertain assumption within abatement cost functions, but it is not comprehensive.

In addition, a more detailed analysis of how the choice of consumption discount rate may affect results and robustness outputs is a natural extension of this work. Discount rates are highly consequential in climate-economy modeling, and are featured prominently in current debates and the literature[56-59]. We incorporate discount rates as a sensitivity (see Methods), while a recent publication incorporated them as one axis of structural uncertainty, concluding with support for a two percent discount rate[32]. We acknowledge the importance of discount rates to this literature, including both the empirical and normative dimensions of the parameter choice, but leave an in-depth discussion and analysis (beyond what is included in the SM) for future research.

Our analysis is silent on the potentially important effects of learning and sequential decision making. We simply derive each candidate policy based on the initial conditions and uncertainties. The policies do not adjust over time even as uncertainty may resolve and policymakers may learn new information. An important related literature uses often sophisticated methods such as dynamic programming or direct policy search to evaluate the impact of uncertainty and/or ambiguity on climate policy[39, 60-63] Our study adopts a simple approach to maintain computational feasibility while including a broad set of structural uncertainties. This simple approach also produces results comparable to optimal policies computed using traditional optimization of expected value. We leave a reframing of this study considering learning for future research.

Policymakers face tradeoffs between spending on abatement and incurring damages from climate change. Applying the minimax regret criterion yields more aggressive abatement, a result driven by the uncertainty in the socioeconomic and damages modules. Minimizing regret hence suggests an aggressive decarbonization policy in response to structural uncertainties. Based on our findings we hypothesize that a particular focus on refining the set of plausible socioeconomic futures and temperature-damage relationships will have the largest impact on these results, as could a more comprehensive approach to abatement costs uncertainty.

The minimax robust policy aligns with the suggestions of some recent single-IAM papers that seek to compute optimal policies under IAMs updated to be consistent with the current literature and compare these policies to international targets[64, 65]. This paper's robust decision-making method arrives at a policy suggestion through consideration of a wide array of IAMs instead of selecting just one. This may make it a more palatable technique for policymakers seeking to utilize IAMs, especially considering the diversity of research and ongoing debates in the field.

Research on different aspects of climate-economy modeling is advancing rapidly, but the uncertainty embedded in modeling future socioeconomic drivers and dynamics, compounded with climatic conditions never-before experienced by humans, makes scientific consensus difficult at best. Decision-making under uncertainty provides policymakers and researchers with a set of tools and criteria designed for this type of deeply uncertain problem with which to evaluate climate policy. This paper combines state-of-the-art research across integrated assessment model components with a decision making under uncertainty regret-based framework to provide useful contributions to discussion about mitigation policy.

## Main References

## Extended data figures

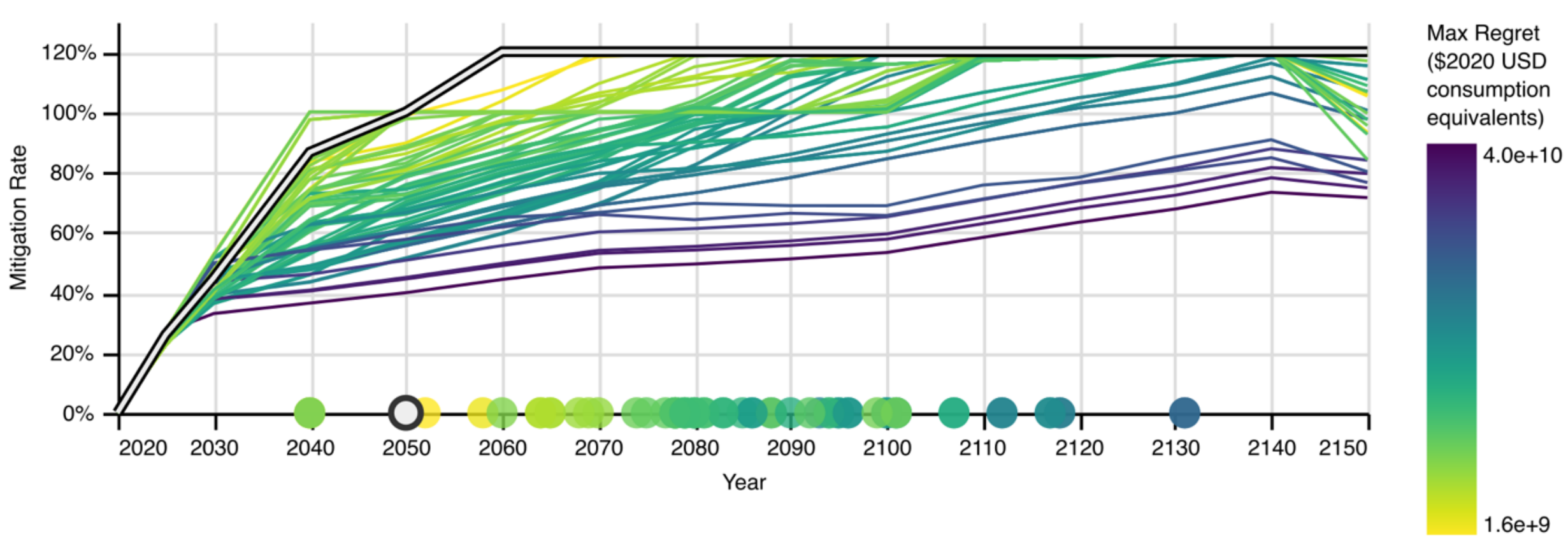

**Extended Data Fig. 1. Optimal mitigation rate paths vary in robustness: the most robust path decarbonizes rapidly.** The minimax regret criterion ranks the robustness of a given policy candidate based on maximum regret (yellow to purple coloring). The white line with a black outline represents the most robust path. Each point on the x-axis shows the date a policy fully decarbonizes date and uses the same coloring scheme.

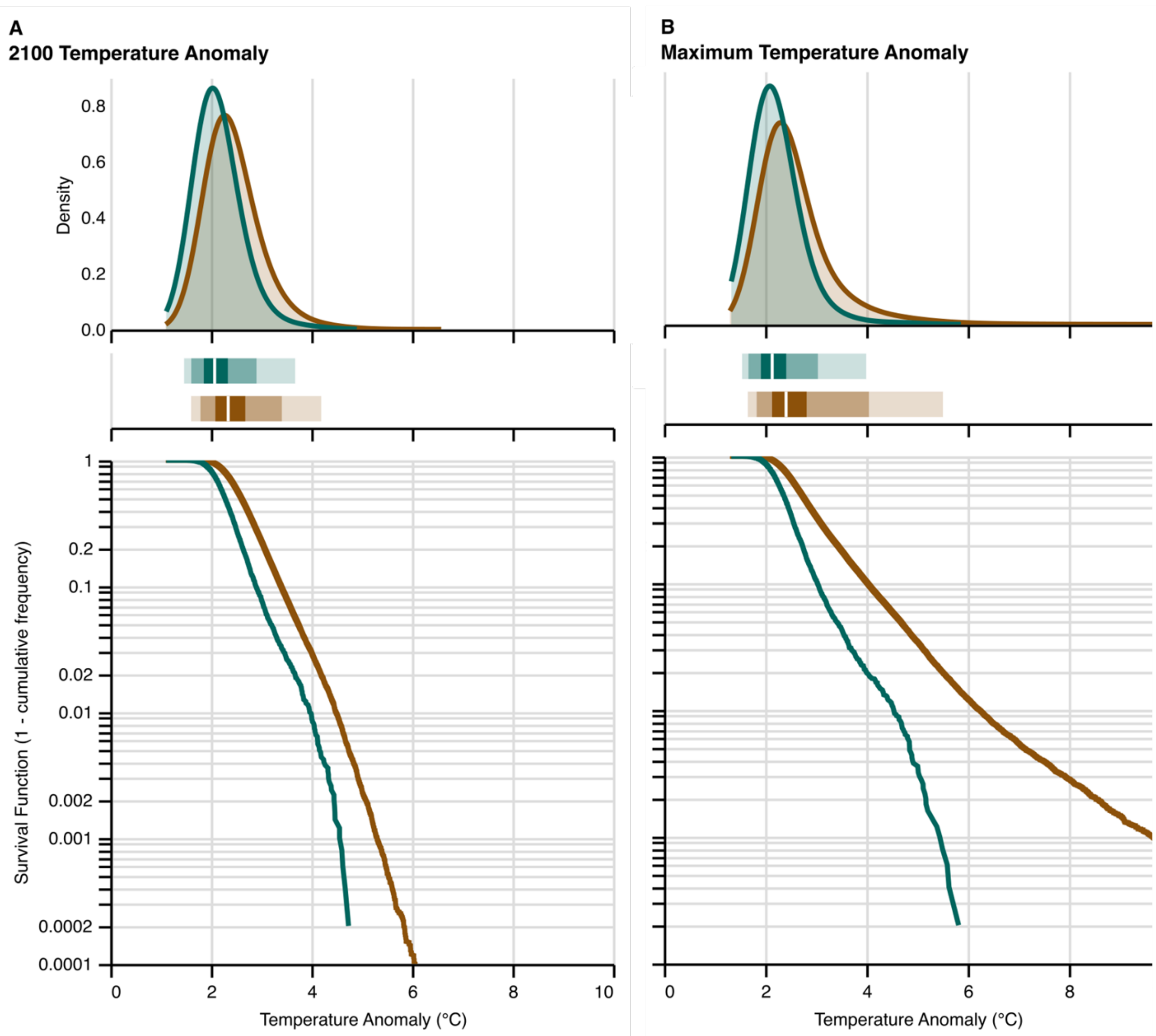

**Extended Data Fig. 2. The robust policy limits temperature increase and avoids worst-case warming scenarios.** The "all policies" PDFs (green) show data points for *n* runs of each policy-state of the world pair, combining the parametric uncertainty of *n* Monte Carlo runs with the structural uncertainty of 100 policies run through 100 potential states of the world. The "most robust policy" display the subset of data points corresponding to using only the most robust policy. Bar plots below the PDFs show the median (white line) and the 25th-75th, 5th-95th, and 1st-99th percentiles with shading of decreasing opacity. The bottom panels show the survival function (1 - the cumulative frequency) in order to focus on the tail behavior of these PDFs We display results for **(A)** the temperature anomaly in 2100 and **(B)** the maximum temperature anomaly across the full time horizon.

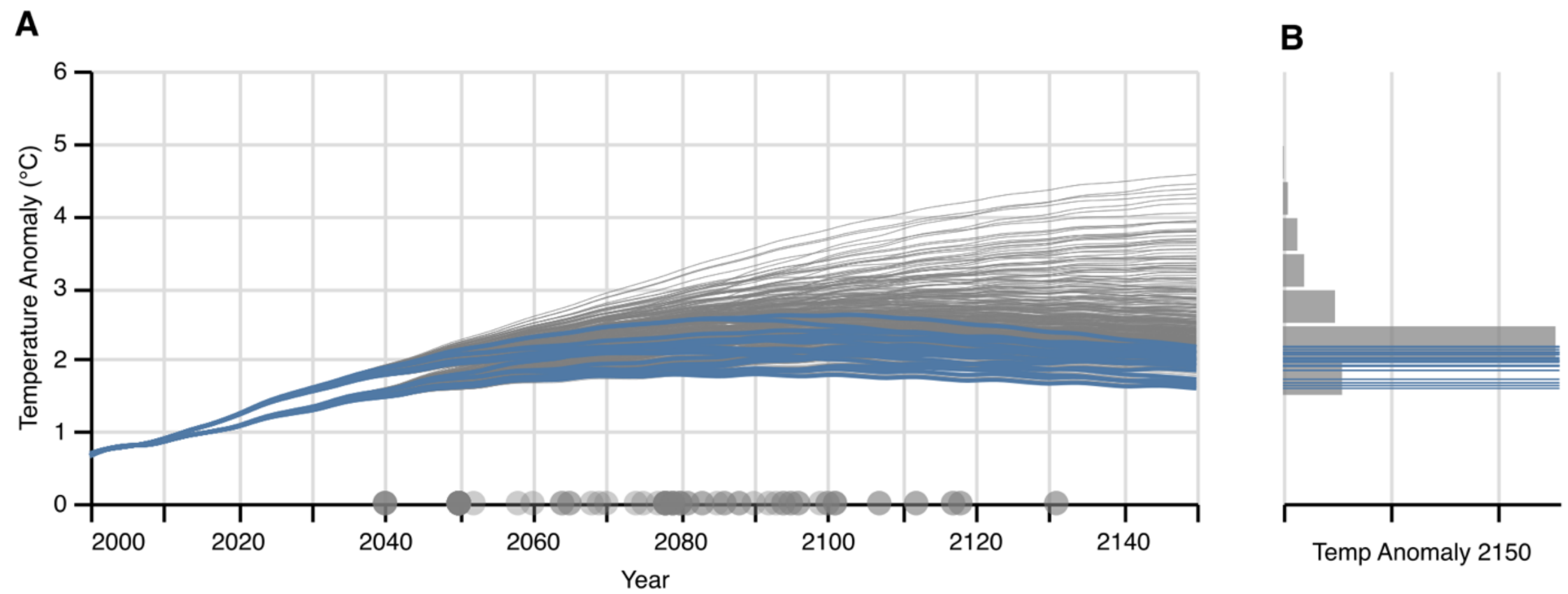

**Extended Data Fig. 3. Robust policy temperature trajectories. (A)** Each policy combined with each state of the world produces a mean temperature anomaly trajectory (grey lines). The most robust policy mitigation path limits the temperature anomaly to the low-end of the range (blue lines). The years of full decarbonization are shown as dots on the x-axis. **(B)** Mean temperature anomalies in 2150 for all policies are shown by the grey histogram, individual blue lines represent mean temperature anomalies for the robust policy under each state of the world.

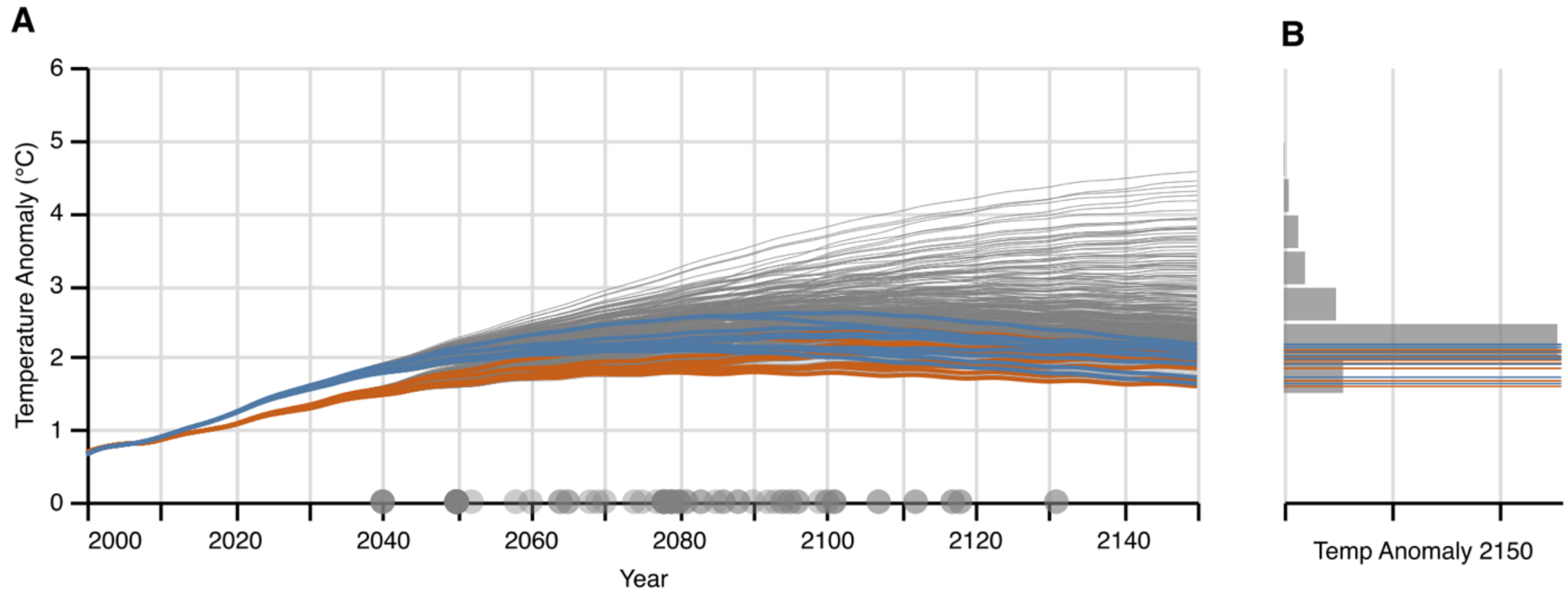

**Extended Data Fig. 4. Robust policy temperature trajectories by climate model. (A)** Each policy combined with each state of the world produces a mean temperature anomaly trajectory (grey lines). The most robust policy mitigation path limits the temperature anomaly to the low-end of the range (blue lines for states of the world including the Finite Amplitude Impulse Response (FaIR) model, orange lines for those including the Simple Nonlinear Earth System (SNEASY) model). The years of full decarbonization are shown as dots on the x-axis. **(B)** Mean temperature anomalies in 2150 for all policies are shown by the grey histogram, individual-colored lines represent mean temperature anomalies for the robust policy under each state of the world.

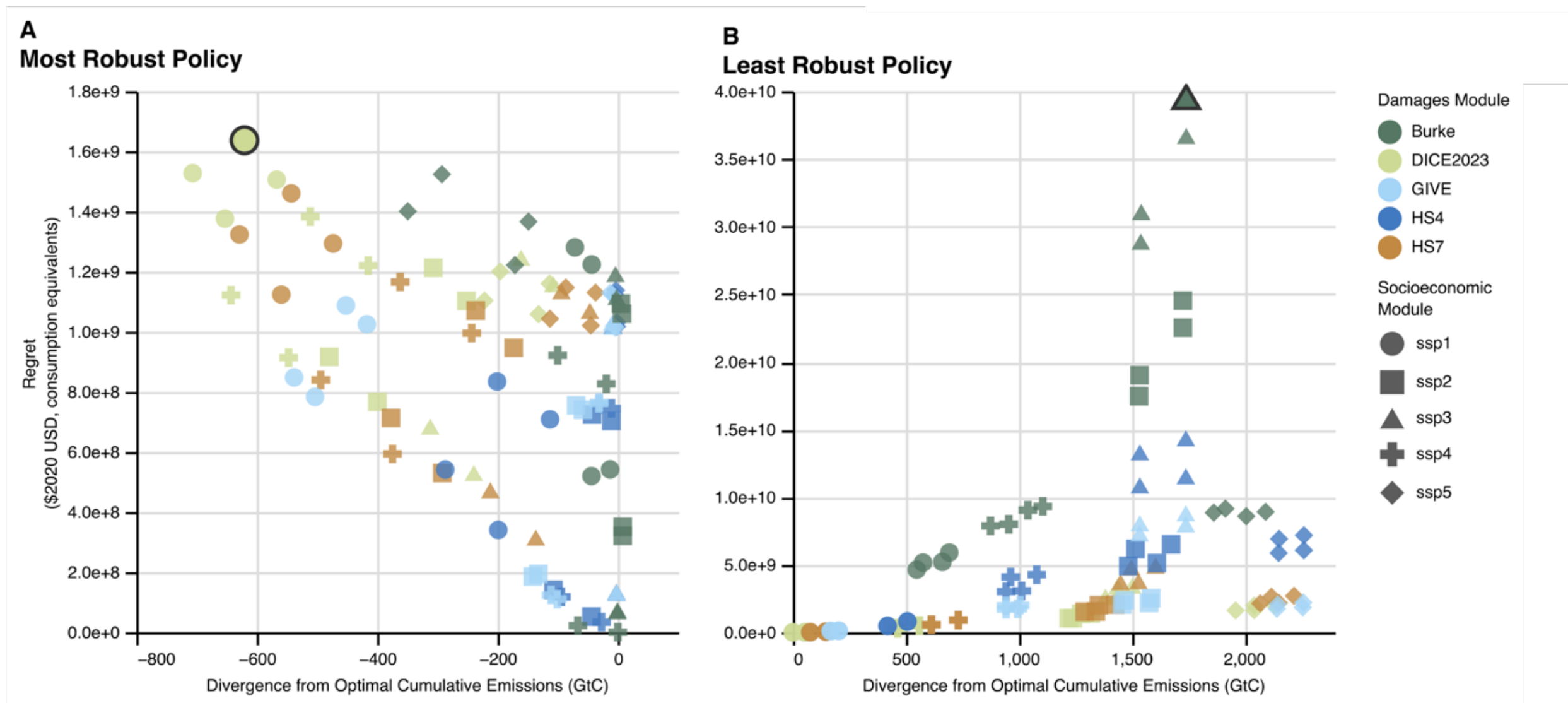

**Extended Data Fig. 5. The socioeconomic and damages modules drive patterns of regret and asymmetry: most and least robust policies.** Each point represents a policy-state of the world (SOW) pair, with the divergence from the optimal cumulative emissions in a realized SOW on the x-axis and the subsequent regret on the y-axis. Colors show the SOW socioeconomic module and shapes show the SOW damages module including Burke et al. (Burke), Dynamic Integrated Model of the Climate and the Economy 2023 (DICE2023), Greenhouse Gas Impact Value Estimator (GIVE), Howard and Sterner specifications 4 and 7 (HS4, HS7). Data points are shown for the **(A)** least robust and **(B)** most robust policy. Larger symbols with a black outline represent the pair that creates maximum regret, one for each policy.

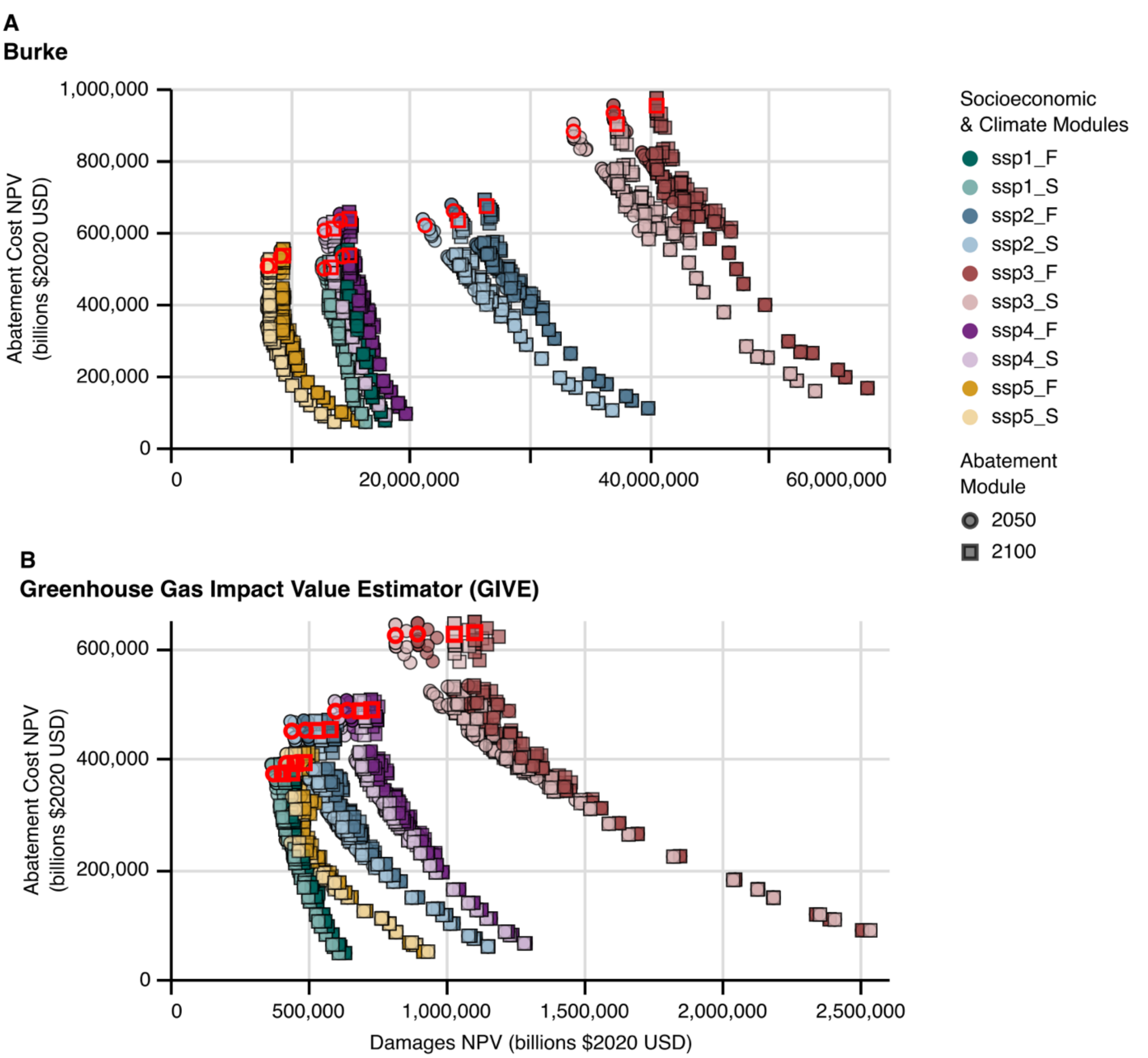

**Extended Data Fig. 6. The minimax regret criterion promotes precaution and relatively high spending on abatement to strictly limit damages from climate change**. Each point represents a policy-state of the world (SOW) pair, with symbols differentiated by characteristics of the state of the world. Colors represent the socioeconomic module, one of five benchmark Shared Socioeconomic Pathways (SSP) and climate module, one of the Finite Amplitude Impulse Response model (F) or the Simple Nonlinear Earth System model (S). Shapes represent abatement costs module, and panels represent two of the damage functions: **(A)** Burke et al. (Burke) and **(B)** Greenhouse Gas Impact Value Estimator (GIVE). This means that each SOW displays 100, each a realization of a candidate policy in that SOW. Bold red outline indicates the most robust policy, showing its abatement costs and damages for that one policy across different future SOW.

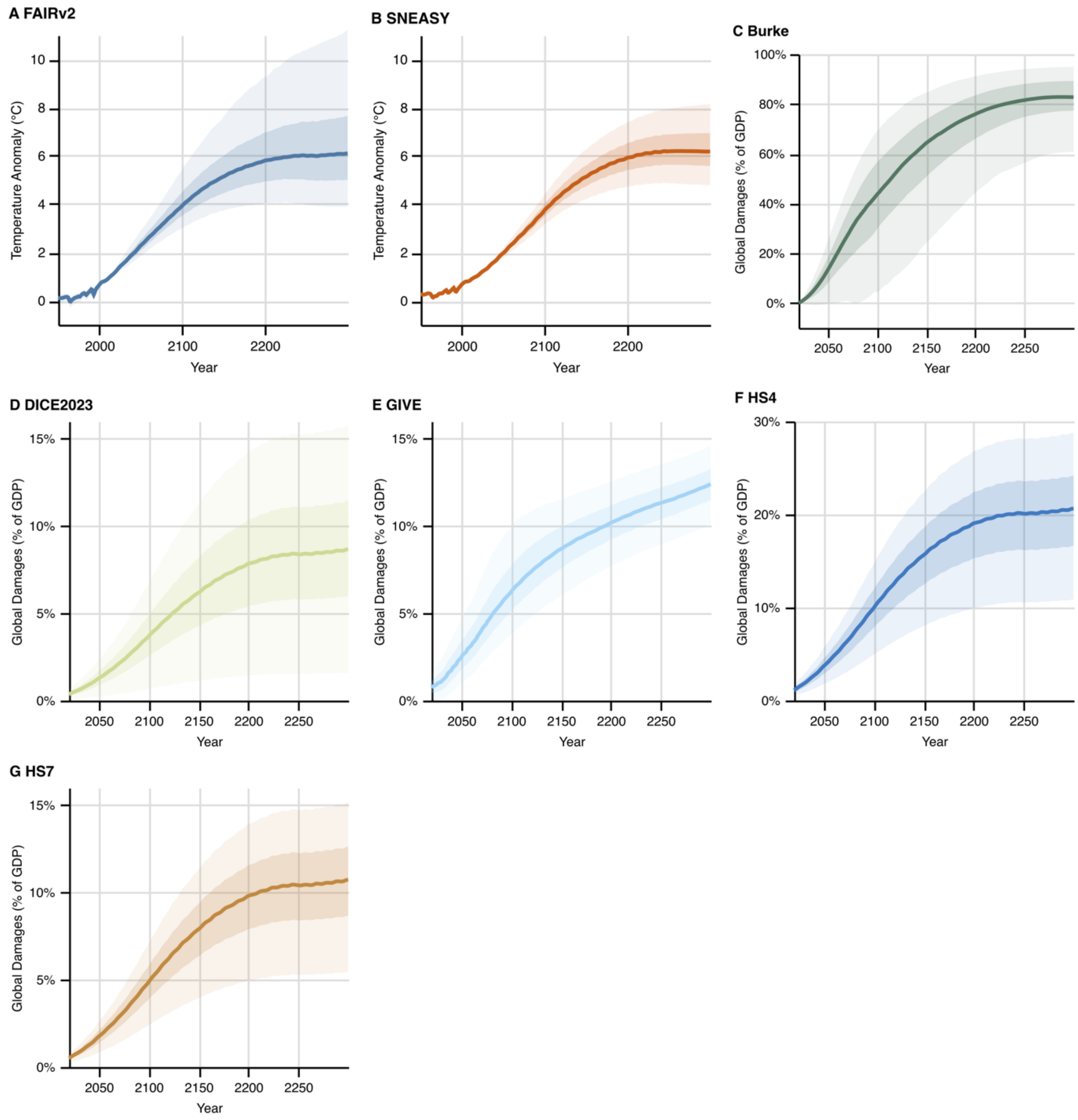

**Extended Data Fig S7. Climate and Damage Module Uncertainty.** Distributions for temperature projections (n = 1,000) for the two climate models (A) Finite Amplitude Impulse Response (FaIR) model and (B) Simple Nonlinear Earth System (SNEASY) model and five damage functions (C) Burke et al. (Burke), (D) Dynamic Integrated Model of the Climate and the Economy 2023 (DICE2023), (E) Greenhouse Gas Impact Value Estimator (GIVE), (F-G) Howard and Sterner specifications 4 and 7 (HS4, HS7). In all panels, solid lines depict the median outcome, with darker shading spanning the 25%–75% quantile range and lighter shading spanning the 5%–95% quantile range. The climate models show temperature anomaly based on the benchmark (no policy) Shared Socioeconomic Projection (SSP) 2 scenario. The damage functions use the same SSP in combination with the FaIR model. Each panel isolates the uncertainty to the parameters within the given module, whereas in our modeling this is compounded with additional uncertainty propagating through the model from other modules.

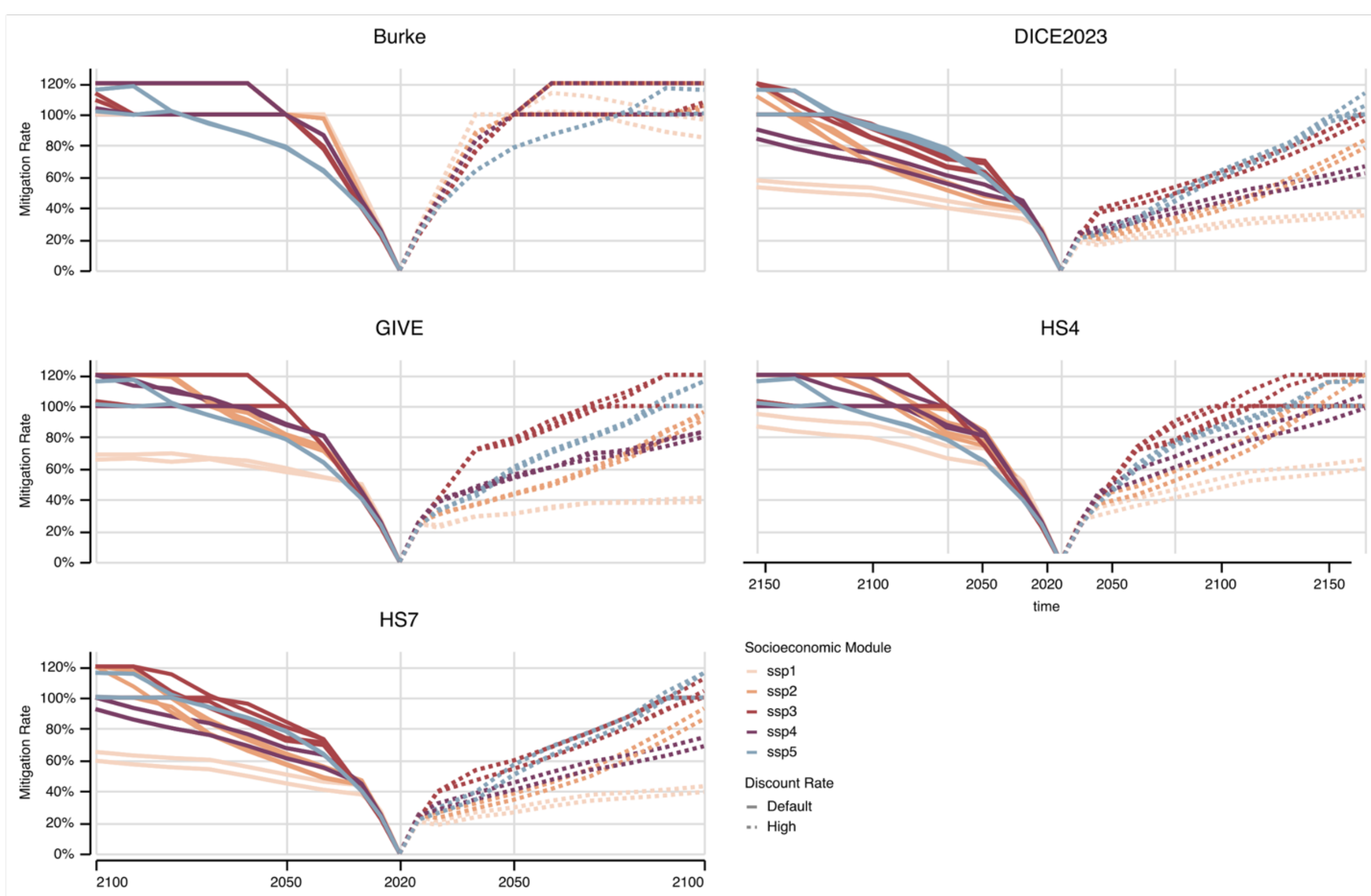

**Extended Data Fig. 8. Mitigation policies for two discount rates.** Comparing default (solid lines) to high (dashed lines) discount rates for all 100 mitigation policies. Colors represent the socioeconomic module, one of five benchmark Shared Socioeconomic Pathways (SSP). Panels represent the damage module which is one the five damage modules Burke et al. (Burke), Dynamic Integrated Model of the Climate and the Economy 2023 (DICE2023), Greenhouse Gas Impact Value Estimator (GIVE), Howard and Sterner specifications 4 and 7 (HS4, HS7)

## Extended data tables

**Extended Data Table 1. Key Terms.** Key terms in this glossary are defined as they are used within this paper.

| Key Term | Definition |
| --- | --- |
| **Model Structure** | Each model structure is composed of four primary modules: socioeconomic, climate, abatement, and damages. Each individual module is based on previously peer-reviewed literature and includes parametric uncertainty whenever possible. This term is used to mechanically describe the composition of an IAM, though it is also the sole source of deep uncertainty and is therefore in some ways synonymous with state of the world. |
| **State of the World** | A future characterized by a unique combination of modules that make up an IAM and encompass only deep, structural uncertainty. A state of the world has numerous uncertain parameters, and thus the combination of a model structure with a particular draw of uncertain parameters encompasses both deep uncertainty and shallow, or parametric uncertainty. |
| **Policy (Candidate)** | A time path of annual carbon dioxide mitigation rates in percentages for the time horizon of the analysis. A policy *candidate* is a policy under consideration in the analysis. |

**Extended Data Table 2. Modules combined to produce new model structures.** Each model structure is composed of four primary modules: socioeconomic (5), climate (2), abatement (2), and damages (6). Each individual module is based on previously peer-reviewed literature and includes parametric uncertainty whenever possible.

| Module Category | Name | Abbreviation | Parametric Uncertainty |
|---|---|---|---|
| Socioeconomic | Shared Socioeconomic Pathway 1 | SSP1 | |
| | Shared Socioeconomic Pathway 2 | SSP2 | |
| | Shared Socioeconomic Pathway 3 | SSP3 | |
| | Shared Socioeconomic Pathway 4 | SSP4 | |
| | Shared Socioeconomic Pathway 5 | SSP5 | |
| Climate | Finite Amplitude Impulse Response model | FaIRv2 | 93,995 samples of uncertain parameters from Leach et al. (2021)[66] for the FaIR model<br><br>10,000 samples of uncertain parameters from Rennert et al. (2022)[8] for BRICK sea-level rise model<br><br>21 unique patterns to recover local mean surface temperatures from global mean surface temperatures from EPA (2023)[100] |
| | Simple Nonlinear Earth System model | SNEASY | 10,000 samples of uncertain parameters for SNEASY and BRICK from Wong et al. (2022a, 2022b) [75, 76] for SNEASY and the BRICK sea-level rise model |
| Abatement | 2016R Dynamic Integrated Climate-Economy model Negative Emissions 2100 | DICE2016R_2100 | Uncertain exponent of cost function and initial cost decline of backstop price distributions based on Lamontagne et al. (2019)[30] and initial backstop price based on Lamontagne et al. (2019)[30] and Butler et al. (2014)[42] |
| | 2016R Dynamic Integrated Climate-Economy model Negative Emissions 2050 | DICE2016R_2050 | |
| Damages | Howard and Sterner Specification 4 | HS4 | |

| Module Category | Name | Abbreviation | Parametric Uncertainty |
|---|---|---|---|
| | Howard and Sterner Specification 7 | HS7 | Uncertain damage coefficient distributions based on Howard and Sterner (2017)[85] |
| | Dynamic Integrated Climate-Economy model (2023) | DICE2023 | Uncertain damage coefficient distribution based on Barrage and Nordhaus (2024)[44] |
| | Greenhouse Gas Impact Value Estimator | GIVE | Uncertain damage coefficient distributions in agriculture and health damage functions based on Rennert at al. (2023)[8] |
| | Burke et al. (2015) | Burke | Uncertain growth effect coefficient distribution in damage function based on 1000 bootstraps of regression based on Burke et al. (2015)[46] |

# Methods

## Model Structures and States of the World

We build 100 model structures based on the existing published literature (Extended Data Table 2) and prioritize covering a range of structures within each of the four primary categories (socioeconomic, abatement costs, climate, and damages) that are compatible with the IAM framework with minimal modifications to their published formulations. We also include parametric uncertainties when available (Extended Data Fig. 7). Each structure characterizes a plausible future state of the world and is used both (1) to create a candidate mitigation policy and (2) as a future state of the world to test the full array of policies against.

As such, we define a state of the world $i$ as the linked combination of four primary modules: socioeconomic module $j$, abatement costs module $k$, climate module $l$, and damage module $m$ that make up one model structure.

$$SOW_i = f(Socioeconomic_j,\ Climate_k,\ AbatementCosts_l,\ Damages_m) \text{ for} \qquad (1)$$

$j$ in SSP1, SSP2, SSP3, SSP4, SSP5,

$k$ in FaIR, SNEASY,

$l$ in DICE2016R_2050, DICE2016R_2100, and

$m$ in Burke, GIVE, DICE2023, HS4, HS7

And

$I = J \times K \times L \times M$ where capital letters refer to number of indices.

*Climate*

We include two climate models.

1. **FaIR**: The Finite Amplitude Impulse Response model (FaIR) is an open-source, emissions-based simple climate model that replicates the relationships between emissions and temperature found in larger Earth system models[66-68]. This model was recommended as a top candidate for use in updated integrated assessment modeling by the National

Academies of Sciences[40] and is used extensively in the literature including in the recent IPCC Sixth Assessment Report (AR6)[67]. Parametric uncertainty is captured through a 93,995 member set of calibrated parameters[66].

2. **SNEASY**: The Simple Nonlinear Earth System Model (SNEASY) is simple climate model with a nonlinear carbon cycle[69, 70]. SNEASY is calibrated extensively with observational data sets. We sample parametric uncertainties using 10,000 sets of parameters from their joint posterior distribution, given the observational data.

These two climate models are calibrated to differing historical datasets, so we normalize the global mean surface temperature anomaly over 2006-2015 relative to a preindustrial period of 1850-1900 to 0.87°C based on IPCC estimates[71].

We use the Building Blocks for Relevant Ice and Climate Knowledge (BRICK) sea-level model to translate global mean surface temperature into global mean sea-level rise[72-74] BRICK estimates global mean sea-level rise as the sum of contributions from the Greenland and Antarctic ice sheets, glaciers and small ice caps, thermal expansion, and land water storage and includes a rapid ice sheet disintegration tipping point representation for the Antarctic ice sheet. We sample parametric uncertainty sets of parameters from Rennert et al. (2022)[8] for coupling with FaIR, and Wong et al. (2022a, 2022b)[75,76] for coupling with SNEASY.

*Socioeconomic*

We use five socioeconomic scenarios, known as the Shared Socioeconomic Pathways (SSPs). For each pathway we obtain a global carbon dioxide emissions trajectory paired with country-resolution population and GDP trajectories and labeled **SSP1-5**[52]. Since we are imposing mitigation policies on top of these paths, we use the benchmark, or "business as usual" (BAU) scenarios from the SSP dataset. We extend these country-resolution projections of GDP and population past the year 2100 to 2300 using a recently published dataset by Benveniste et al. (2020)[77]. Guided by the literature, we ramp $CO_2$ emissions down linearly to zero by 2250[78].

We recognize recent important work in this domain including incorporating expert elicitation to create probabilistic socioeconomic projections[47] and addressing endogenous sociological, political, and technological processes[51], we are limited to using BAU policies that do not incorporate assumptions about future mitigation policy layered on top of current policy.

*Abatement Costs*

Both abatement costs modules are based on the 2016R Dynamic Integrated Model of Climate and the Economy (DICE2016R), one of a series of the DICE model[79]. We chose to use the DICE2016R model here, as opposed to DICE2023, for two reasons. First, DICE2023's abatement costs function is modified to include non-$CO_2$ greenhouse gases, and is calibrated as such, so it is inconsistent with this paper's application to only $CO_2$. Second, we prioritize the ability to include uncertainty around parameters, which we do not have for DICE2023's abatement costs function.

The feasibility of negative carbon emissions, both in terms of timing and magnitude, is consequential for choosing mitigation rate policies[3, 80-82]. Deep uncertainty about technological processes, costs, and their interplay with sociopolitical factors make it difficult to make predictions about negative emission[83, 84]. We introduce this element of uncertainty by creating two abatement costs structures. The more optimistic structure allows for negative emissions after 2050 mirroring the use of negative emissions technologies by about mid-century in recent IPCC publications[3]. The second delays the feasibility of negative emissions until after 2100. These two modules are labeled **DICE2016_2050** and **DICE2016_2100** respectively. The maximum feasible mitigation rate is set to 120% of baseline emissions rate as implemented in DICE. Further constraints on mitigation rates are explained in the *Generating Candidate Policies* section below. We note that negative carbon emissions feasibility and timing is just one of several important uncertainties in abatement costs, yet we are constrained by the availability of abatement cost functions compatible with our analytical framework. This is an area for future improvements to this work.

We represent parametric uncertainty using Uniform distributions for the exponent of the control cost function[30], the initial backstop price[42], and the initial cost decline backstop cost per period[30] (Extended Data Table 2).

*Damages*

We implement five damage functions that translate changes in mean surface temperature to impact on the economy, or more specifically to GDP. Four of these functions connect change in temperature to impact on the *level* of GDP in a given year, while the other two represent damage as impacting GDP *growth***.**

1. **HS4 and HS7:** Howard and Sterner (2017) present a meta-analysis-based global damage function employing a quadratic function of mean global temperature anomaly to produce damages as a fraction of annual global GDP[85]. We include the authors' preferred regression specification 4 (HS4) as well as specification 7 (HS7), as used in recent rulemaking analysis by the Environmental Protection Agency[85, 86]. Table 2 in the main Howard and Sterner (2017) publication provides the parameterizations and uncertainty specifications for this damage function.

2. **DICE2023:** The 2023 Dynamic Integrated Model of Climate and the Economy (DICE2023) uses a meta-analysis-based global damage function, the latest in a series of DICE damage functions commonly used in the literature[44]. This quadratic function of temperature anomaly generates impacts as a fraction of annual GDP. We include parametric uncertainty on the quadratic term coefficient.

3. **GIVE:** The Greenhouse Gas Impact Value Estimator (GIVE) is an open-source integrated assessment model that comprehensively responds to the near-term recommendations of the 2017 National Academies of Sciences, Engineering, and Medicine report[8. 40]. Individual components in GIVE are based on separate peer-reviewed publications, including four damage categories that collectively make up the GIVE impact estimates: agriculture[87], mortality[88], energy consumption[89], and sea-level rise[90]. The agriculture and mortality functions contain parametric uncertainty as fully described in the main publication. For computational efficiency we replace the sea-level rise damages component of the GIVE model with a reduced form version as generated and used in a recent paper by Nordhaus[91]. The net impact of these four sectoral damage functions represents the levels-based damage estimates of the GIVE model at a spatial resolution of 184 individual countries.

4. **Burke:** Burke et al. (2015)[46] presents our second empirically driven growth-effects damage function. This widely-cited paper estimates a *non-linear*, concave function relating economic productivity to temperature, identifying peak productivity around 13°C, below which temperature increases have positive effects, and above which they have negative effects. This model computes a growth effect term to be applied to the counterfactual annual GDP growth without a temperature change. We use the authors preferred specification, the pooled response (does not differentiate between low- and high-income countries) estimating the short-run, 0-lag effect. We use a bootstrapped distribution with 10,000 draws to represent uncertainty on the growth effect coefficient.

Extending damage functions to estimate impacts over several centuries, in our case through 2300, can pose analytical challenges, and their original publications often limit analysis to the 21st century. This is especially true for damage functions that estimate persistent impacts on GDP. For example, in Burke et al. (2015)[46] the unbounded benefits to cold countries from high levels of warming accumulate to unintuitively and unreasonably high values by the 22nd century, dominating the catastrophic damages to hotter countries, which are intrinsically bounded at the baseline GDP levels[46]. For the countries with the coldest baseline temperatures such as Mongolia, Iceland, and Finland, benefits from warming can reach thousands of times baseline GDP levels within the time horizon.

Following recent literature tackling this same challenge in an optimization framework, we limit impacts for growth-impact damage specifications to +500% (bound on benefits) -100% (bound on damages) of baseline GDP in any given country-year, noting the damages bound of 100% is by construction in the as-published damage functions[90, 91]. This loosens the restrictions imposed by Gazzotti et al. (2021, 2022)[92, 93], which use [+100%, -100%) and allows impacts within the century to mirror those in the original Burke et al. (2015)[46] implementation yet bounds results to prevent extreme behavior.

## Generating Candidate Policies

We define a policy candidate as a time path of $CO_2$ mitigation rates from 2020 to 2300. More specifically, we create a choice variable at 5–10-year increments from 2025 to 2150 and hold mitigation rates constant from 2150 to 2300. We create 100 policy candidates by optimizing a social welfare function to come up with the $CO_2$ mitigation pathway that maximizes expected welfare. To retain computational feasibility, we produce one policy candidate for each state of the world (SOW), taking the expectation across all *n* parameterizations of the same SOW to evaluate the performance of a vector of mitigation rates under a particular model structure. We generate a policy $x$ by maximizing expected welfare across all $n$ parameterizations of state of the world $s$

$$E[w_s] \tag{2}$$

where a policy $x$ is defined as a vector of mitigation rates subject to a set of constraints listed below, the expectation is taken across all $n$ parameterizations of state of the world $s$, and $w_s$ is the welfare derived from applying the mitigation rates in SOW $s$ under one of $n$ parameterizations. We employ a social welfare function $w$ using a discounted utility framework, as is most common for climate economics literature and policy[94-96], as follows.

$$w = \sum_t L_t \left(\frac{1}{(1+\rho)^t}\right) U(c_t) \tag{3}$$

Where $L_t$ is population in year $t$, $c_t$ is average per capita consumption in year $t$ and the parameters $\rho$ and $\eta$ represent the pure rate of time preference and elasticity of marginal utility of consumption, or colloquially risk aversion, respectively. The utility function $U$ is defined as

$$U(c) = \begin{cases} \ln c\,, & \eta = 1 \\ \frac{c^{1-\eta}}{1-\eta}\,, & \eta \neq 1 \end{cases} \tag{4}$$

We choose the central specifications for these parameters using a recently published expert elicitation and set $\rho$ to 0.5% and $\eta$ to 1.0[97]. A long and ongoing literature in climate economics and public policy debate the choice of discounting parameterizations, especially under uncertainty and over long time-horizons[56-59]. These parameterizations are not a structural uncertainty in this paper and, instead, are varied for sensitivity analysis (see Sensitivity Analysis section) but have been incorporated into similar frameworks in the literature[32, 34].

The mitigation rate in year $t$ is bounded between 0% and 100% through the year of negative emissions (2050 or 2100), after which it has a maximum of 120% as described above in the *Abatement Costs* section. Computing optimal policies using the DICE abatement costs function can result in unrealistic mitigation policy scenarios which mitigate more, or faster, than the literature suggests is possible[30, 66, 98]. We mirror the approach used in van der Wijst et al. (2021)[98] and introduce bounds guided by the recent IPCC Special Report of Global Warming of 1.5°C pathway[3, 98]. In short these include (1) a minimum of -20 $GtCO_2$ per year, which may be a stricter constraint than 120% of baseline emissions in certain socioeconomic scenarios (2) a maximum reduction rate of 2.2 $GtCO_2$ per year.

**Evaluate Policy Performance in Each State of the World**

We next evaluate the performance of each policy candidate in each state of the world and assign each candidate-SOW pair a measure of performance based on the expected welfare across all parameter sets for a SOW, producing 100 x 100 performance values. Again, for simplicity of presentation and to reduce computational complexity, we do not include the performance of all uniquely parameterized SOWs, but aggregate performance based on the expected welfare across all parameterized SOWs characterized by a given model structure.

**Choosing the Most Robust Policy**

To choose the most robust policy we first calculate the regret for each policy-SOW pair, and then use the minimizing maximum (minimax) regret criterion to choose the policy candidate with the lowest maximum regret across all model structures. We choose the minimax regret criterion for its history of use in decision theory literature[12, 43] and its fairly wide application in environmental and climate domains[29, 32-34]. We also consider an alternative metric minimizing the mean (minimean) regret.

The regret for a policy-SOW combination is the difference between how the policy performs in the SOW, and how the optimal policy for that SOW performs. The most robust policy under the minimax regret criterion is the one with the smallest maximum regret across the 100 values. Note the intuition here is that a policy paired with the same SOW it was optimized for will produce a regret metric of 0. The regret for a given policy $x$ and state of the world $s$ is calculated informally as the difference in welfare between the result of implementing policy $x$ and the best we can possibly do in realized state of the world $s$, or more formally,

$$r_{x,s} = w_{x,s} - w_{v_{optimal},s} \tag{5}$$

Where $r$ is regret, $x$ is our candidate policy defined as a vector of mitigation rates, $v_{optimal}$ is the optimal policy in state-of-the-world $s$, $w$ is welfare as defined in equation 3. Note again that if $x = v_{optimal}$ then regret will be zero.

**Alternative Policy Candidates (Fig. 2)**

Figure 2 displays alternative policy candidates compared to the most robust policy. Below we detail the generation of these policies and any caveats relevant to comparisons to current literature.

1. DICE 2023: We generate the DICE2023 path by combining (1) Shared Socioeconomic Pathway 2 (2) the FaIR climate model (3) the DICE2023 damage function and (4) the DICE2016 abatement costs function allowing negative emissions in 2100. This is an approximation that matches the original DICE2023 model as closely as possible without introducing inconsistencies with the present framework.

2. GIVE: We generate the DICE2023 path by combining (1) Shared Socioeconomic Pathway 2 (2) the FaIR climate model (3) the GIVE damage function and (4) the DICE2023 abatement costs function allowing negative emissions in 2050. This is an approximation that matches the original GIVE model as closely as possible without introducing inconsistencies with the present framework, noting that the original GIVE model does not include an abatement costs function.

3. Expected value over structural uncertainty: This policy is generated through a single optimization in which the objective function averages over performance in all possible states of the world, effectively implementing the Laplace Principle of Insufficient Reason and assuming all model structures are equally likely. Importantly note that this policy is not one of the set of 100 policies considered in the rest of the analysis, but an additional policy for comparison. We maximize the expected welfare for each state of the world $i$, each of which is the expectation across all $n$ parameterizations as described in equation 3.

$$x = \frac{1}{100}\sum_{i=1}^{100} E\left[w_{s_i}\right] \qquad (6)$$

4. Minimizing mean regret: This is the policy which has the minimum mean regret – as opposed to our primary robustness metric which minimizes maximum regret – across the states of the world.

## Sensitivity Analysis

*Discount Rates*

The consumption discount rate employed in climate-economy modeling has been the topic of considerable debate and discussion for decades[56, 59]. It is consequential for economic evaluation of climate policy, especially given the long time horizon over which the calculations of climate impacts must be made and the substantial uncertainty in climate damages and economic growth[57, 58]. In this paper we choose one discounting specification for the main analysis, and include a second, higher rate as a sensitivity:

- Central: $\rho = 0.5\%$ and $\eta = 1.0$ from Drupp et al. (2018)[97]
- High: $\rho = 1.0\%$ and $\eta = 1.5$ from Barrage and Nordhaus (2024)[44]

A third addition could be a low discount rate, for example using $\rho = 0.1\%$ and $\eta = 1.0$ as suggested by Stern (2006)[99].

Altering the discount rate used in the analysis affects the two steps that employ social welfare function: (1) calculating optimal mitigation policy for each state of the world by maximizing expected welfare and (2) calculating expected welfare from implementing each policy in each state of the world.

As expected from the literature, a higher discount rate generally results in a less stringent policy (Extended Data Fig. 8). The impact of changing the discount rate on the optimal path for a given policy varies. For example, using a high discount rate may have little effect on policy paths that are bounded by the feasibility limits placed on mitigation path, but a larger visible effect shows for those with more moderate damages. We hypothesize that with a sufficiently high discount rate even highly pessimistic path (eg. those characterized by high impact damage functions high business as usual emissions) would become significantly less stringent.

The primary findings of a precautionary, stringent robust pathway and asymmetry of regret are robust to the specification of discount rate. That said, we hypothesize that at the limit a sufficiently high discount rate may reduce the asymmetry by strongly discounting climate damages and reducing the tail of damages.

*Different Robustness Metrics*

There are a wide variety of available robustness metrics in the literature, and different metrics will indicate different rankings of policy alternatives[45]. McPhail et al. (2018)[45] provide a valuable overview of these metrics and their consequences, separating them into expected value metrics, metrics of higher-order moments, regret-based metrics, and satisficing metrics. In this paper we focus on the regret-based subset, which highlight the downside risk and help us to answer the question of "what if the model we choose is wrong?". Our central results use the minimax regret criterion, ranking each policy based on the maximum regret incurred by that policy. We also consider minimizing the mean regret (minimean), ranking each policy based on the average regret it incurs. Qualitatively these two metrics produce similar results through 2050 and then diverge on timing of net negative emissions. Both are stringent mitigation policies relative to the considered set. Recent literature also shows that regret-based metrics tend to produce the same rankings of policies[23, 45]. Examining other categories of robustness metrics, such as including satisficing metrics, may alter rankings and results more significantly[31].

*Pattern-Scaled Climate Variables*

We recover local temperatures by pattern-scaling the global mean surface temperature outputs of the FaIR model using patterns derived from 21 global circulation models[100]. Details on this approach can be found in the documentation of the Github repository for this reference.

## Methods References

**Acknowledgements and disclaimers**

This work is not a product of the United States Government or the United States Environmental Protection Agency. The authors are not doing this work in any governmental capacity. The views expressed are those of the authors and do not necessarily represent those of the United States Government or the United States Environmental Protection Agency.

**Author contributions**

** initials in each category are listed in paper author order*

Conceptualization: L.R., D.A.

Methodology: L.R., D.A., K.K.

Investigation: L.R., F.E., D.S., B.P, K.K., D.A.

Visualization: L.R., F.E., D.S., B.P., K.K., D.A.

Funding acquisition:

Project administration: D.A.

Supervision: K.K., D.A,

Writing – original draft: L.R.

Writing – review & editing: L.R., F.E., D.S., B.P., K.K., D.A.

**Competing interest declaration**

Authors declare that they have no competing interests.